\documentclass[amssymb,amsmath,pra,showpacs,twocolumn,floatfix,superscriptaddress]{revtex4}
\usepackage{graphicx}
\usepackage{lastpage}
\usepackage{amssymb}

\begin{document}

\title{High speed quantum gates with cavity quantum electrodynamics}

\author{Chun-Hsu Su}
\email{chsu@ph.unimelb.edu.au}
\affiliation{Quantum Communications Victoria, School of Physics, University of Melbourne, VIC 3010, Australia}

\author{Andrew D. Greentree}
\affiliation{Quantum Communications Victoria, School of Physics, University of Melbourne, VIC 3010, Australia}

\author{William J. Munro}
\affiliation{Hewlett-Packard Laboratories, Filton Road, Stoke Gifford, Bristol BS34 8QZ, United Kingdom}
\affiliation{National Institute of Informatics, 2-1-2 Hitotsubashi, Chiyoda-ku, Tokyo 101-8430, Japan}

\author{Kae Nemoto}
\affiliation{National Institute of Informatics, 2-1-2 Hitotsubashi, Chiyoda-ku, Tokyo 101-8430, Japan}

\author{Lloyd C.L. Hollenberg}
\affiliation{Quantum Communications Victoria, School of Physics, University of Melbourne, VIC 3010, Australia}

\date{\today}

\begin{abstract}
Cavity quantum electrodynamic schemes for quantum gates are amongst the earliest quantum computing proposals. Despite continued progress
and the recent demonstration of photon blockade, there are still issues with optimal coupling and gate operation involving high-quality cavities.  Here we show that dynamic cavity control allows for scalable cavity-QED based quantum gates using the full cavity bandwidth.  This technique allows an order of magnitude increase in operating speed, and two orders reduction in cavity $Q$, over passive systems. Our method exploits Stark shift based $Q$ switching, and is ideally suited to solid-state integrated optical approaches to quantum computing.
\end{abstract}


\maketitle

\section{Introduction}
Quantum information technology enables new forms of communication and computation that are more efficient than existing classical approaches. Applications as diverse as secure communication~\cite{bennett84ekert91}, simulating quantum systems~\cite{feynman82}, and factoring~\cite{shor97} have already been identified, and experimental techniques essential to realize these and other applications are advancing steadily. Recent, notable results include those based on optical~\cite{kok07} and trapped-ion systems~\cite{schmidt03chiaverini05}, and architectures have been proposed that incorporate quantum error correction and address various issues associated with scalability~\cite{kielpinski02}.

Amongst the variety of physical systems being explored, cavity quantum electrodynamics (QED) systems have shown great promise. A high quality ($Q$-factor), small volume ($V$) optical cavity represents an almost ideal environment for achieving coherent atomic manipulation at the single-quantum level with minimal dissipation. Such high-$Q/V$ cavities allow for deterministic atom-photon coupling in a near or far-off resonant configuration that elicit linear or Kerr-like nonlinear atomic response. These processes are central to some of the pioneering proposals -- single photon sources~\cite{law97}, quantum bus protocols~\cite{spiller06}, quantum computation and communication schemes~\cite{pellizzari95duan01} -- and demonstrations of quantum-phase gate~\cite{turchette95}, quantum memory~\cite{maitre97} and photon blockade~\cite{birnhaum05dayan08}. Cavity-assisted photonic networks for preparing 2D and 3D cluster states have also been discussed~\cite{devitt07stephens08devitt08}.

\begin{figure}[tb!]
\includegraphics[width=1.03\columnwidth,clip]{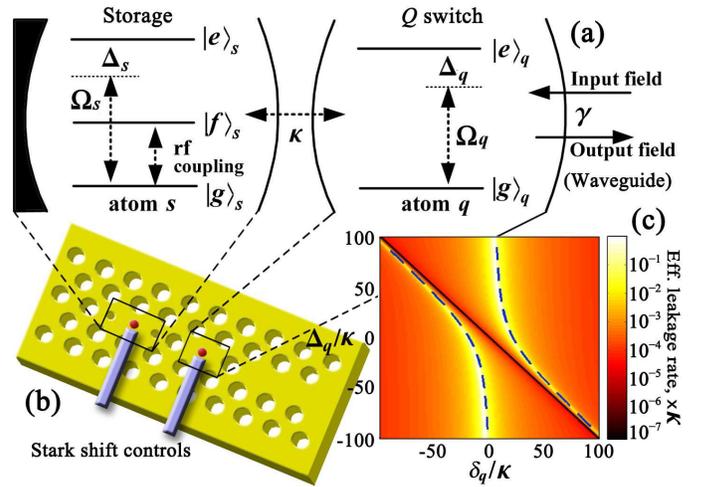}
\caption{\label{fig:schem} (color online)
(a) Schematic of proposed CZ gate. Atom-$s$ is the target qubit with qubit transition $|g\rangle_s\leftrightarrow|f\rangle_s$ addressed via an RF field. The atom accumulates a $\pi$ phase shift conditioned on the populated cavity mode $s$, which dispersively couples the $|g\rangle_s$ to $|e\rangle_s$ transition. The atoms are coupled to their respective cavity mode with Rabi frequency $\Omega_\alpha$ and detuned by $\Delta_\alpha$ ($\alpha = s, q$). Spectral tuning of atom-$q$ allows dynamic and coherent photon switching between the waveguide and the storage cavity. The cavities are coupled with photon-hopping rate $\kappa$, detuned by $\delta_q$, and $\gamma$ is the cavity decay rate. (b) Proposed gate design in PBG lattice, missing holes constitute the cavities and waveguide. Red circles denote atoms tuned via e.g. Stark shift gates (rectangles). (c) Light leakage rate from mode $s$ in the static two-cavity arrangement. Cavity loss is rapid at storage-switch resonance (dashed) and is minimized at two-photon resonance ($\Delta_q = -\delta_q$) between mode $s$ and atom $q$.}
\end{figure}

Despite these promising applications, the fundamental time-bandwidth relation of passive high-$Q$ cavities limits operational speeds and ultimately practicality. Worse still, oscillations in the photon intensity due to poorly matched pulses are serious source of error. Here we show that the use of appropriate dynamic control breaks these limitations and reduces required cavity $Q$ by two orders of magnitude compared with passive schemes, and at the same time realizes high-fidelity faster quantum gate.

The dynamic control approach used here modifies the cavity-waveguide coupling, from high $Q$ during confinement to low $Q$ during in/out-coupling, with a tuning time much shorter than the photon lifetime of the cavity~\cite{xu07}. This approach can also be exploited for active pulse shaping. Although $Q$ switching is a standard practice for classical lasers, there are few schemes~\cite{greentree06, bermel06fernee07} that work at the quantum level and are compatible with solid-state cavities such as photonic-band-gap (PBG) structures.

Within this context, we report the first theoretical study of a controlled phase (CZ) gate between photonic (control) and matter-based (target) qubits using active $Q$ switching. The CZ gate is a two-qubit entangling gate which induces a phase flip if the control and the target are in a specified state, i.e. for two arbitrary qubits, it performs the transformation $|11\rangle\rightarrow-|11\rangle$. Specifically, the state of an atom inside a cavity is controlled by the photonic occupation of the cavity via usual Jaynes-Cummings interaction in the strong coupling and dispersive regime. Using a two-cavity setup, we show that quantum interference allows high reflectivity and optimal photonic confinement of the interaction cavity. Through an adiabatic process controllable by the switch, ultra low and relatively fast transfer of excitations between the interaction cavity and waveguide is possible. To generate a scalable architecture for quantum computing, this form of CZ gate is integral to coupling between matter-based qubits where stationary-to-flying qubit interconversion is used. Our scheme is a powerful enabler in preparing cluster states through operator measurement across multiple photonic qubits~\cite{devitt07stephens08devitt08} and for integrated optics at the quantum level. 
This scheme is discussed generically, but is pertinent for strong coupling systems, e.g. semiconductor quantum dots in photonic crystal cavities~\cite{englund07} and circuit-QED~\cite{schuster07bishop08}. PBG structures in diamond are also developed to exploit the vast resource of optically active color centers for such applications~\cite{greentree08}.

\section{Implementing a controlled phase gate}
The proposed CZ gate is shown schematically in Fig.~\ref{fig:schem}(a). To realize the $Q$ switch, a two-level atom $q$ is coupled to the cavity mode $q$, detuned by $\Delta_q$, with single photon Rabi frequency $\Omega_q$. In the storage cavity, the target qubit (atom-$s$) is a three level system with ground $|g\rangle_s$, metastable $|f\rangle_s$ and excited $|e\rangle_s$ states. $|g\rangle_s\leftrightarrow|f\rangle_s$ is the qubit transition that can be controlled via resonant RF coupling. To achieve a linear gate, we operate in the dispersive regime where the $|g\rangle_s\leftrightarrow|e\rangle_s$ transition couples to the cavity mode $s$ with Rabi frequency $\Omega_s \ll \Delta_s$. This interaction is preferred over more complicated resonant schemes which suffer from increased sensitivity to timing noise. To illustrate CZ operation, we consider, without loss of generality, the starting states $|\pm\rangle \equiv (|f\rangle_s\pm|g\rangle_s)/\sqrt{2}$, and the gate is described by the unitary operator $\mathcal{U} \approx {\rm exp}[-i(\Omega_s^2t/\Delta_s)|1\rangle_{ss}\langle 1|\otimes|g\rangle_{ss}\langle g|]$  where $|1\rangle_s$ denotes a photon in mode $s$. After time $T_{\pi} \equiv \pi\Delta_s/\Omega_s^2$, a phase flip is induced on $|g\rangle_s$
so that $|\pm\rangle_s \rightarrow |\mp\rangle_s$.

Before continuing with the construction of the gate, we note that without switching, i.e. only the coupled atom-cavity system on the left of Fig.~\ref{fig:schem}(a), we need a cavity of low bandwidth and strong atom-cavity coupling to effect a phase-flip via scattering. To illustrate our points, we consider an implementation in the optical regime with nitrogen-vacancy (NV) diamond defect center in a PBG cavity fabricated in diamond. The center is placed at the maximum of the cavity mode with wavelength near the zero-phonon line resonance of the center $\lambda = 638$~nm and frequency 2.95~PHz. This corresponds to the transition from the excited spin triplet state ($^3E$) to the $m=0$ sublevel of the triplet ground state ($^3A$). Assuming that the cavity is sub-micrometer in dimensions with $V\sim\lambda^3$, then $\Omega_s \approx 10$~GHz~\cite{greentree06}. In this regime, the phonon sidebands will be suppressed~\cite{su08}, justifying the three-state approximation of the center where $|e\rangle_\alpha\equiv |^3E,m=0\rangle, |g\rangle_\alpha \equiv |^3A,0\rangle$ and $|f\rangle_s\equiv|^3A,\pm1\rangle$ for $\alpha = s, q$. A 2.88~GHz RF field allows a complete control over its ground state transitions~\cite{howard06}. To effect a gate under these conditions implies a 0.1~$\mu$s pulse and a technically-challenging static $Q$ of $10^8$ for a gate error rate of $10^{-3}$. When $Q=10^6$ is used, the photon only succeeds in inducing $\sim0.14\pi$-phase shift. In contrast, active switching can achieve a gate fidelity of 0.993 with nanosecond photon pulse, nanosecond gate time and more modest $Q\sim\mathcal{O}(10^5-10^6)$. This fidelity is sufficient to implement the topological error correction scheme described in Ref.~\cite{raussendorf07} using the cluster state preparation network introduced in Ref.~\cite{devitt07stephens08devitt08}. Cavities in this range have been demonstrated in silicon~\cite{noda07}. Recently, cavity modes in diamond photonic crystal cavities near 638~nm with $Q = 585$ has also been demonstrated~\cite{wang07} and there exist suitable diamond cavity designs~\cite{bayn07tom06} to achieve the required $Q/V$. To implement schemes which demand higher fidelity gates a higher $Q/V$ cavity will be required. 

The coupled-cavity system, in the dipole and rotating-wave approximations, is governed by the Hamiltonian,
\begin{eqnarray}
\mathcal{H} & = & \Delta_s|e\rangle_{ss}\langle e| + \Big(\delta_q - i\frac{\gamma_q}{2}\Big)a_q^\dagger a_q + \Big(\delta_q + \Delta_q - i\frac{\gamma_e}{2}\Big) \nonumber\\
  & \times & |e\rangle_{qq}\langle e| + \int_{-\infty}^{\infty}\omega b^\dagger(\omega)b(\omega)d\omega + \Big(\Omega_s\sigma^{+}_sa_s + \kappa a_s^\dagger a_q \nonumber \\
  & + & \Omega_q \sigma^{+}_q a_q + \int_{-\infty}^{\infty} \sqrt{\frac{\gamma}{2\pi}}b^\dagger(\omega)a_q d\omega + {\rm h.c.} \Big)
	\label{eq:ham}
\end{eqnarray}
where $\sigma^{+}_\alpha\equiv|e\rangle_{\alpha\alpha}\langle g|$ is the atomic raising operator for atom-$\alpha$, and 
$a_\alpha$ ($a_\alpha^\dagger$) is the annihilation (creation) operator of the cavity. $b(\omega)$ [$b^\dagger(\omega)$] is its counterpart for a photon of frequency $\omega$ in the waveguide, and satisfy the Heisenberg equation of motion~\cite{walls95},
\begin{equation}
	\dot{b}(\omega) = -i\omega b(\omega) + \sqrt{\frac{\gamma}{2\pi}} a_q.
\end{equation}
where $\gamma \equiv \omega_c/(2 Q)$ is the cavity decay rate, where $\omega_c$ is the cavity resonant frequency. The resonant frequency of the cavities differs by $\delta_q$. In the tight-binding regime, the cavities are coupled with photon-hopping rate $\kappa$. 

To treat decoherence, we introduce $\gamma_e$, the spontaneous emission rate of atom $q$ and $\gamma_q$, the transverse cavity decay rate. Decoherence of the atomic qubit becomes negligible in the dispersive regime as the emission rate from $|e\rangle_s$ scales with $(\Omega_s/\Delta_s)^2$. It is implicit in our treatment that transverse decay rate of the left cavity is weak compared to the required confinement time $T_{\pi}$.

In the one-quantum manifold, from $|\dot{\psi}\rangle = -i\mathcal{H}|\psi\rangle$, it is straightforward to obtain a set of differential equations for the probability amplitudes ($C^\xi_\alpha, D^\xi_\alpha$) for states, namely $|\xi,1\rangle_s|g,0\rangle_q|{\rm vac}\rangle$ ($C_s^\xi$), $|e,0\rangle_s|g,0\rangle_q|{\rm vac}\rangle$ ($D_s$), $|\xi,0\rangle_s|g,1\rangle_q|{\rm vac}\rangle$ ($C_q^\xi$), $|\xi,0\rangle_s|e,0\rangle_q|{\rm vac}\rangle$ ($D_q^\xi$), and $|\xi,0\rangle_s|g,0\rangle_q|\phi\rangle$ ($C_{\rm out}^\xi$), where $\xi$ identifies the dressed basis of atom-$s$ and the third ket indicates the photonic state in the waveguide. In turn, the amplitude of the output pulse $f_{\rm out}^\xi$ is related to the input ($f_{\rm in}^\xi$) by the standard input-output relation $f_{\rm out}^\xi = \sqrt{\gamma} C_q^\xi - f_{\rm in}^\xi$~\cite{walls95}. These equations can be solved for the system dynamics numerically. To optimally couple the photon to the cavity and perform a gate requires three steps. These are: 

\textit{Step A} -- \textit{Loading}. Near-resonance coupling of atom $q$ with mode$q$ is best expressed in the well-known dressed basis $|\pm\rangle_q$~\cite{walls95}. Since the eigenenergies vary with atomic frequency, the system can be made transmissive (at resonance) to an incoming pulse on the waveguide. Sweeping the switch through this resonance by tuning $\Delta_q$ permits mode-matched transfer of a left-travelling photon into mode $s$. This occurs when we evolve the joint storage-switch system along a particular energy eigenstate that has a standard form for a dressed $\Lambda$-atom system~\cite{shore90},
\begin{equation}
	|\Phi\rangle = \frac{1}{\mathcal{N}}\Big\{\kappa \Omega_q|1\rangle_s + \mathcal{E}\Omega_q|1\rangle_q + [\mathcal{E}(\mathcal{E}-\delta_q)-\kappa^2]|e\rangle_q\Big\}
\end{equation}
normalized with $\mathcal{N}$, where $\mathcal{E}$ is its eigenenergy.
In particular, $|\Phi\rangle\approx|1\rangle_s$ when the storage is weakly coupled to the switch for some $\Delta_q = \Delta_q^{\rm off}$, and $|\Phi\rangle\approx|1\rangle_q$ when mode $s$ is resonant with $|\pm\rangle_q$ for $\Delta_q = \Delta_q^{\rm on} \equiv (-\delta_q^2+\Omega_q^2)/\delta_q$. Thus as we spectrally tune atom $q$, we effect coherent transfer from the waveguide to mode $s$ via the switch. The evolution requires adiabaticity parameterized by,
\begin{equation}
	\mathcal{A} \equiv \frac{|\langle\Phi'|\dot{\mathcal{H}}|\Phi\rangle|}{ |\langle\Phi'|\mathcal{H}|\Phi'\rangle-\langle\Phi|\mathcal{H}|\Phi\rangle|^2} \ll 1.
	\label{eq:adia}
\end{equation}
where $|\Phi'\rangle$ is the eigenstate closest to $|\Phi\rangle$ in energy.
The position of storage-switch resonance is set up with detuning $\delta_q$. The use of large $|\delta_q|$ lengthens the switching time $T_{\rm sw}$ as the matrix element for photon hopping $_{q}\langle \pm|\kappa a_q^\dagger a_s|1\rangle_s$ weakens, whereas small $|\delta_q|$ implies that atom $q$ must be tuned over an increasing range as off- and on-resonance points become farther apart [Fig.~\ref{fig:schem}(c)].

\textit{Step B} -- \textit{Gate}. Once the photon is inside the cavity, we decouple mode $s$ from the waveguide for a duration of $T_{\pi}$ to enact $|+\rangle_s\rightarrow|-\rangle_s$. High gate fidelity $\mathcal{F}\equiv |C_{\rm out}^-(\infty)|^2$ is achievable, as we show, when $T_{\rm sw} \ll T_{\pi}$. However, one can improve the gate time and fidelity with an additional control to perform an adiabatic modulation of dispersive interaction, where the rate of phase shift is controlled via tuning of atom-$s$. The idea is to use a suitable $\Delta_s$ to induce rapid phase flip over (shorter) $T_{\pi}$, and a large $\Delta_s$ during switching improves the fidelity by avoiding over-rotation. We now further develop the underlying physics for light confinement and identify the ideal value for $\Delta_q^{\rm off}$.

The coupled-cavity solution of our setup is a spatial analogue of electromagnetically-induced transparency in a $\Lambda$-system~\cite{zhou08}. At exact two-photon resonance between mode $s$ and atom $q$ for $\Delta_q = \Delta_q^{\rm off}\equiv -\delta_q$, $|\Phi\rangle$ is maximally decoupled from $|1\rangle_q$, and a favoured population in $|1\rangle_s$ when $\Omega_q \gg \kappa$. In this configuration, the storage-waveguide coupling is minimized, offering both optimal photonic confinement and reflectivity to any incoming light [Fig.~\ref{fig:schem}(c)]. Although some finite overlap with $|e\rangle_q$ leads to leakage, this probability scales with $(\kappa/\Omega_q)^2$.

\textit{Step C} --  \textit{Unloading}. The photon is removed from the cavity using a time-reversed Stark tuning. Being the time-reversal of \textit{Step A}, this final step restores the initial photon and leaves the atom in its desired state. 

\begin{figure}[tb!]
\includegraphics[width=0.9\columnwidth,clip]{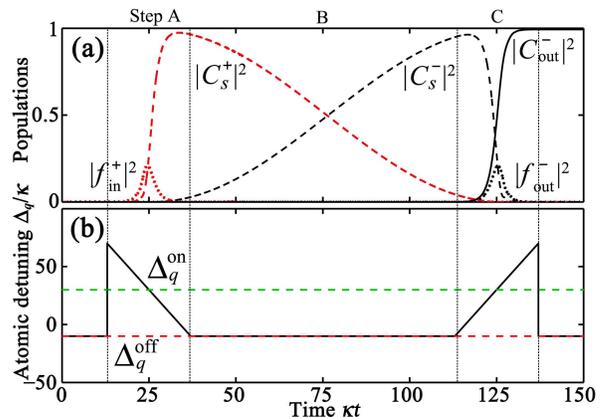}
\caption{\label{fig:phmod} (color online) (a) Simulation of the CZ gate. The matter qubit (atom-$s$) is initialized in $|+\rangle_s$. \textit{Step A}: A single photon of shape $|f^{+}_{\rm in}|^2$ (red dotted) is switched into the storage cavity adiabatically ($|C_s^+|^2$, red dashed). \textit{Step B}: Dispersive interaction results a phase shift from $|+\rangle_s$ to $|-\rangle_s$ (black dashed). \textit{Step C}: The photon is returned to the waveguide. The gate fidelity $\mathcal{F}\equiv|C_{\rm out}^-|^2=0.993$ (black solid). $|f_{\rm out}^-|^2$ (black dotted) is the pulse envelope of the output photon. (b) Corresponding atomic detuning $\Delta_q(t)$ in unit of $\kappa$, that varies between $\Delta_q^{\rm off}$ where the switch has minimal transmittivity and $\Delta_q^{\rm on}$ where $|1\rangle_s$ is resonant with $|-\rangle_q$. Parameters are $\kappa=\gamma$, $\Omega_s=5\kappa$, $\Omega_q=20\kappa$, $\delta_q = 10\kappa$, $\Delta_s=10^3\kappa$, and $\gamma_q, \gamma_e=0$.}
\end{figure}

\section{Results and Discussion}
With above formalism, we simulate the operation of the CZ gate with realistic parameters in Fig.~\ref{fig:phmod}. An input single-photon pulse is switched into the storage using a linear shift $\Delta_q$ over a time scale of $10\kappa^{-1}$. In this case ($\mathcal{A}\sim10^{-3}$), the success probability for adiabatic transfer $P_{\rm out} = 1- \mathcal{O}(10^{-4})$. While the optimal input amplitude has been constructed as the complex conjugate time-reversal of a switched photon, it is a Gaussian pulse easily prepared with a conventional coherent source. Such pulses are standard for cavity-QED gate systems, being ideal for fiber transmission and Hong-Ou-Mandel interferometry~\cite{rohde05}. When the waveguide is decoupled, the leakage error is $10^{-3}$ for the choice $\Omega_q = 20\kappa$. The likelihood of single-photon absorption by atom-$s$ must be $\eta \ll 1$. Here since $\Delta_s \approx \Omega_s/\sqrt{\eta}$ for $\eta = \mathcal{O}(10^{-5})$, it is an unlikely source of error. Immediately after the photon is removed, the qubit is phase-flipped with $\mathcal{F}=0.993$. Most importantly, we have demonstrated that the storage time is much longer than both the photon lifetime of the stand-alone device and the time-bandwidth of the travelling photon.

\begin{figure}[tb!]
\includegraphics[width=\columnwidth,clip]{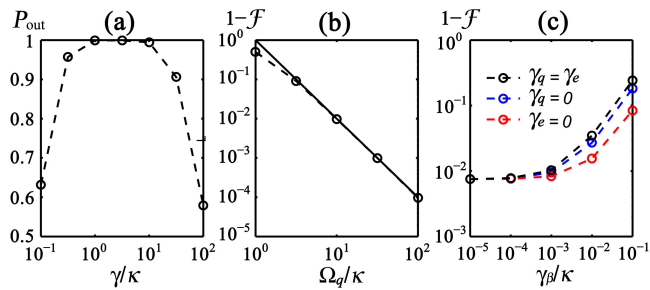}
\caption{\label{fig:errors} (color online) Circles denote data points from simulations. Dashed lines are guides for the eye. (a) Success probability for light outcoupling via adiabatic transfer for fixed $T_{\rm sw}=10/\kappa$, as a function of cavity decay rate $\gamma/\kappa$. (b) Reduction in gate fidelity $\mathcal{F}$ due to premature leakage, as a function of atom-cavity coupling $\Omega_q/\kappa$. The solid curve plots $(\kappa/\Omega_q)^2$, showing a good overlap. (c) $\mathcal{F}$ is calculated for different decoherence rates $\gamma_\beta$ ($\beta = q, e$) for the scenario in Fig.~\ref{fig:phmod}. The black curve denotes the case where the rates are equal, while other curves distinguish individual contributions. 
Parameters follow from Fig.~\ref{fig:phmod} unless stated otherwise.}
\end{figure}

Proceeding with further simulations, we first ignore the decoherence and study the effect of the cavity decay rate $\gamma$ on success probability ($P_{\rm out}$) for light transfer from mode $s$ to waveguide. Fig.~\ref{fig:errors}(a) shows that this probability peaks at $\gamma/\kappa = \mathcal{O}(1)$ for a fixed switching time. This is contrary to the expectation that a large $\gamma/\kappa \gg 1$ should suggest the light field would be dumped immediately from mode $q$ after the adiabatic transfer. 

Thus far, leakage before switching is an inherent error due to state mismatch between $|\Phi\rangle$ and $|1\rangle_s$. In Fig.~\ref{fig:errors}(b), we show $\mathcal{F}$ versus $\Omega_q/\kappa$ where the error rates due to imperfect adiabatic transfer and absorptions are explicitly excluded. It proves that the overlap $_q\langle e|\Phi\rangle$ leads to this error, and thus can be suppressed by increasing $\Omega_q$ with higher $Q/V$ cavities. For instance, $\mathcal{F}\sim0.9999$ is feasible when $\Omega_q/\kappa = 100, \Delta_s = 10^4\kappa$ and decoherence is suppressed accordingly. At last, we take decoherence at the switch into consideration. In Fig.~\ref{fig:errors}(c), we expect that the fidelity is fundamentally limited by premature leakage, and degrades as the decoherence rate $>10^{-2}\kappa$. 

When realized with NV centers ($\gamma_e = 10$~MHz) in PBG cavities of $Q = 10^6$, this gate operates with $\sim$20~ns pulse and a gate time of 200~ns for $\mathcal{F} = 0.989$. However, an extra control $\Delta_s(t)$ realizes $\mathcal{F} = 0.996$ and a gate time $\sim2T_{\rm sw}=$ 40~ns. Stark tuning of isolated centers via an external control field has been demonstrated~\cite{tamarat06}. The tuning range from the early work of Redman \textit{et al.} of order 1~THz~\cite{redman92} is more than enough for the proposed scheme. In the microwave regime, when superconducting qubits ($\gamma_e = 1$~MHz, $\omega_c \sim 1$~GHz) are coupled to stripline cavities with rate $0.1$~GHz~\cite{schuster07bishop08}, $Q\sim10^2$ suffices for a similar fidelity with $2~\mu$s pulse and $\sim4~\mu$s gate time. 

\section{Conclusion}
Advances in fabrication have led to the development of ultra-small, low loss, solid-state cavities, with obvious potential for quantum information applications.  However passive devices suffer from the unavoidable time-bandwidth relation that severely limits their performance. Active $Q$ switching breaks this nexus and allows the full bandwidth of the cavities be used. We have shown that dynamic Stark-shifting with coupled cavities permits high speed single-photon $Q$ switching, realizing an order of magnitude faster two-qubit (CZ) gate with less stringent $Q$ requirement. This is a significant step in improving the prospects for solid-state cavity-QED based quantum logic, and motivates the further experimental effort in coupled-cavity QED.

\section*{ACKNOWLEDGMENTS}
We thank A.~M.~Stephens, Z.~W.~E.~Evans, C.~D.~Hill and S.~J.~Devitt for valuable discussions. WJM and KN acknowledge the support of QAP, MEXT, NICT and HP. CHS, ADG and LCLH acknowledge the support of Quantum Communications Victoria, funded by the Victorian Science, Technology and Innovation (STI) initiative, the Australian Research Council (ARC), and the International Science Linkages program. ADG and LCLH acknowledge the ARC for financial support (Projects No. DP0880466 and No. DP0770715, respectively).

\clearpage
\newpage
\onecolumngrid

\end{document}